\documentclass[journal]{IEEEtran}
\usepackage{cite} %生成参考文献
\usepackage{amsmath} %多行公式支持1
\usepackage{amsfonts,amssymb} %空心字母支持
\usepackage{mathrsfs} %花体字母支持 \mathscr{A} \mathcal
\usepackage{graphicx,epsfig,balance} %插入图片支持
\usepackage[utf8]{inputenc} %使用UTF8编码，注意CJK也要从GBK改为UTF8
\usepackage{CJK} %中文支持
\usepackage{multirow}
\usepackage{bm} %加粗希腊字母\bm{}
\usepackage{algorithm} %algorithm框图支持
\usepackage{algorithmic}
\usepackage{extarrows} %长等号支持
\usepackage{color} %颜色支持
\usepackage{ragged2e} %两边对齐
\usepackage{caption}
\usepackage{subfigure}

\newcommand{\tu}{\textup}

\begin{document}

\title{Deep Reinforcement Learning for UAV Navigation Through Massive MIMO Technique}

\author{Hongji Huang, Yuchun Yang, Hong Wang, Zhiguo Ding, Hikmet Sari, and Fumiyuki Adachi
\thanks{H. Huang, H. Wang, and H. Sari are with Key Lab of Broadband Wireless Communication and Sensor Network Technology, Ministry of Education, Nanjing University of Posts and Telecommunications, Nanjing 210003, China. H. Sari is also with Sequans Communications, 92700 Colombes, France. (E-mail: \mbox{hongji.huang@ieee.org}, wanghong@njupt.edu.cn, hsari@ieee.org).}
\thanks{Y. Yang is with Jilin University, Changchun 130012, China. (E-mail: yuchunyang.ee@gmail.com)}
\thanks{Z. Ding is with School of Electrical and Electronic Engineering, University of Manchester, Manchester, M13 9PL, U.K. (E-mail: zhiguo.ding@manchester.ac.uk)}
% \thanks{S. Guo is with Department of Computing, The Hong Kong Polytechnic University, Kowloon, Hong Kong. (E-mail: song.guo@polyu.edu.hk)}
\thanks{F. Adachi is with Wireless Signal Processing Research Group, Research Organization of Electrical Communication (ROEC), Tohoku University, Sendai 980-8577, Japan. (E-mail: adachi@ecei.tohoku.ac.jp)}
% <-this % stops a space
}

\maketitle
\begin{abstract}
Unmanned aerial vehicles (UAVs) technique has been recognized as a promising solution in future wireless connectivity from the sky, and UAV navigation is one of the most significant open research problems, which has attracted wide interest in the research community. However, the current UAV navigation schemes are unable to capture the UAV motion and select the best UAV-ground links in real-time, and these weaknesses overwhelm the UAV navigation performance. To tackle these fundamental limitations, in this paper, we merge the state-of-the-art deep reinforcement learning with the UAV navigation through massive multiple input multiple output (MIMO) technique. To be specific, we carefully design a deep Q-network (DQN) for optimizing the UAV navigation by selecting the optimal policy, and then we propose a learning mechanism for processing the DQN. The DQN is trained so that the agent is capable of making decisions based on the received signal strengths for navigating the UAVs with the aid of the powerful Q-learning. Simulation results are provided to corroborate the superiority of the proposed schemes in terms of the coverage and convergence compared with that of other schemes.
\end{abstract}

\begin{IEEEkeywords}
Massive multiple input multiple output (MIMO), deep reinforcement learning, UAV navigation
\end{IEEEkeywords}

% \fi
\IEEEpeerreviewmaketitle

\section{Introduction}

Future communication networks must tackle not only incremental throughput and explosive traffic, but also for energy consumption, ultra-reliability, and supporting highly diversified applications with heterogeneous quality-of service (QoS) requirements \cite{mag}. Therefore, wireless connectivity techniques have drawn universal attention in academy and industry communities, and several emerging techniques have been proposed, such as balloons \cite{bal}, and unmanned aerial vehicles (UAVs) \cite{uav1}. Particularly, thanks to their wide applications, high mobility, and superior line-of-sight (LoS) propagation, UAVs have a great potential to be the airborne nodes such as relays and terminals, and therefore they are considered as an essential part of future networks.

In the recent years, a large quantity of works have been devoted to enhancing the performance of the UAV-enabled communications. In \cite{z1}, the authors proposed a UAV-enabled data collection system for optimizing the energy consumption issue, where a UAV is assigned to collect data from a ground terminal at the fixed location. To deal with the endurance problem, a novel scheme which leverages the proactive caching at the users was provided and numerical results have demonstrated that proactive caching is a good candidate to resolve the endurance issue in the UAV-based systems \cite{z2}. Taking the advantages of the massive multiple input multiple output (MIMO) technique which can boost the system capacity, a UAV cellular-based system through massive MIMO was explored and this approach can increase the reception reliability \cite{cemi}.

Due to the high mobility of the UAVs, UAV navigation is an important technique and it has already been applied in the public safety, emergency rescue, and search operation. In \cite{ev1}, an evolutionary-based scheme was proposed which integrates the classic genetic algorithm into a breeder genetic algorithm, but this method cannot realize a reliable UAV navigation because of the randomness of the genetic algorithm and its complicated implementation. With the aid of sensors, the authors presented an autonomous UAV navigation approach (TF-UAV) to address the simultaneous localization and mapping issues but it requires robots which hinder its flexibility \cite{tfu}. When we use the received signal strength indicator (RSSI) for UAV navigation, the occurrence of deep fades degrades the performance if the emerging methods are not introduced. Also, UAVs always move in a wide range and it is of great significance to make them work automatically for improving their communication coverage. But the conventional algorithms cannot satisfy the increasing coverage because UAV movements require much energy and slow navigation deteriorates the coverage performance. Recently, by exploiting the potentials of machine learning into wireless communication, a deep learning-based wireless communication method provides an alternative mean for optimizing the UAV navigation problem, whose performance has been corroborated in non-orthogonal multiple access (NOMA) \cite{h1}, massive MIMO \cite{h2,h0}, traffic control \cite{NK2017,nk2}, routing techniques \cite{nk3}, software defined network (SDN) \cite{ne}, UAV \cite{uav,uav6}, and millimeter-wave (mmWave) communication \cite{h4}, etc.. In particular, \cite{uav6} proposed a deep learning-based method for UAV navigation without requiring sensing data that provides mapping information, however this method cannot converge quickly, which makes it difficult to be applied in real-time navigation scenarios. Reinforcement learning. which is a branch of machine learning that can address model-free problems by leveraging past observation and rewards, is a state-of-the-art method to produce control policies by using action space. It should be pointed out that the number of the action spaces is determined by the complexity of the state. In 2015, Deepmind proposed a reinforcement learning-based framework called deep Q-network (DQN) \cite{nature}, which integrates the deep learning into the Q-learning. DQN is a promising tool to address multi-agents optimization problems such as the UAV navigation.

Inspired by the above considerations, in this paper, we incorporate the deep reinforcement learning technique into the UAV navigation through the massive MIMO. The main contributions of this paper are listed as follows.
\begin{enumerate}
  \item First, we employ the deep reinforcement learning technique to achieve UAV navigation through the massive MIMO. By constructing a DQN \cite{nature}, we obtain the optimal location selection policy based on the received signal strengths. Different from the previous works which mainly introduce the speed or geographic position for UAV navigation, the proposed method converges quickly and it also realizes good coverage performance.
  \item Second, based on the developed DQN, we propose an efficient deep learning-based scheme for optimizing UAV navigation performance. After training the DQN, an environment simulator is developed for UAV navigation with a better coverage and faster convergence. Furthermore, extensive numerical results are provided to verify the superior performance of the proposed DQN navigation schemes.
\end{enumerate}

\section{System Model}
\label{sec2}

Consider a special massive MIMO system, which comprises one mobile BS with $N_t$ antennas and $K$ UAVs with single antenna. According to the well-known ray-tracing-based wireless channel model, the channel model of the $k$-th UAV at the $t$-th time slot is formulated as
\begin{align}
\label{h1}
\mathbf{h}_k(t) = \int_{\varphi_{\tu{min}}}^{\varphi_{\tu{max}}} \mathbf{a}^t(\varphi_{k}(t)) g_k^t(\varphi_{k}(t)) d\varphi_{k}(t),
\end{align}
where $\mathbf{a}^t(\varphi_{k}(t))$ and $g_k^t(\varphi_{k}(t))$ represent the array response and the complex gain coefficient of the $k$-th UAV at time slot $t$, respectively. To be specific, $\varphi_{k}(t)$ denotes the incidence angle of the $k$-th UAV, and $\varphi_{\tu{min}}$ and $\varphi_{\tu{max}}$ are its minimum and maximum values, respectively. Here, $\mathbf{a}^t(\varphi_{k}(t))$ can be written as
\begin{align}
\mathbf{a}^t(\varphi_{k}(t)) = \frac{1}{\sqrt{N_t}} [1, e^{\frac{-j2\pi d}{\lambda} \sin{\varphi_{k}(t)}},\cdot \cdot \cdot, e^{\frac{-j2\pi d}{\lambda} (N_t - 1) \sin{\varphi_{k}(t)}}]^T,
\end{align}
Here, it is noted that $d$ is the antenna size, while $\lambda$ represents the carrier wavelength. The autocorrelation of the gain coefficient of the $k$-th UAV can be expressed as
\begin{align}
\label{h0}
\mathbb{E}\{g_k^t(\varphi_{k}(t)) g_k^{\ast,t}(\varphi_{k}(t^{'}))\} = \gamma^k \upsilon_k^t(\varphi_{k}(t)) \delta(\varphi_{k}(t) - \varphi_{k}(t^{'})),
\end{align}
where $\gamma^k$ represents the received signal power at UAV $k$, and $\upsilon_k^t(\varphi_{k}(t))$ is the power azimuth spectrum at time slot $t$ that describes the power distribution of the channel in the angle domain. It is noted that $\gamma^k$ often experiences severe deep fades, leading to serious errors in UAV navigation. Therefore, we provide a deep reinforcement learning-based scheme to address this issue for boosting the UAV navigation performance. Besides, $\delta(\cdot)$ is denoted as the Dirac delta function and it has the property that $\int_{-\infty}^{+\infty} \delta(\varphi) d\varphi = 1$. Furthermore, the signal-to-interference-plus-noise ratio (SINR) $\eta_k$ at the $k$-th UAV is formulated as
\begin{align}
\label{h2}
\eta_k(t) = \frac{P_{\tu{tr}} |\mathbf{h}_k^H(t) \mathbf{w}_k|^2}{P_{\tu{tr}} \sum\limits_{j\neq k, 1 \leq j \leq K}|\mathbf{h}_k^H(t) \mathbf{w}_j|^2 + \sigma_k^2},
\end{align}
where $P_{\tu{tr}}$ is the transmitted power at the BS, and $\mathbf{w}_k$ is the unit-norm vector for the $k$-th UAV. Also, we assume that this system is corrupted by additive white Gaussian noise (AWGN) with zero mean and variance $\sigma_k^2$. The UAV navigation is based on the RSSI and $\eta_k(t)$ is used to compute the immediate reward as described in the next section. We consider the maximum communication range of each UAV is $R$, and the coverage range $R_c$ is defined as the communication range for the ground users when the UAV flies in the sky with $R_c \leq R$.

\section{DQN-based UAV Navigation Framework}

In this section, we provide a deep reinforcement learning-based framework for UAV navigation through the massive MIMO. To be specific, we first develop a DQN framework, and then formulate a learning policy to train the developed network. Furthermore, we propose an efficient deep reinforcement learning-based strategy for UAV navigation.

\subsection{Deep Q-network}

As an embranchment of machine learning, reinforcement learning has attracted great attention among academia and industry. For attaining the best situation, multi-agents interact with the environment and they search for the optimal strategy with the maximum reward. Generally speaking, reinforcement learning can be regarded as a specific description of Markov decision processes (MDPs). It is comprised of four elements: a policy, a reward signal, an environment, and a utility function, which is a good candidate for resolving the high-complexity situations and capturing the realistic scenarios.

However, the conventional reinforcement learning requires the agents to adopt the appropriate representations of the environment based on the high-dimensional input and generate past knowledge to the new state. Meanwhile, its applicability only covers the low-dimensional area where the features can be fully exploited. To break up these gaps, DQN which integrates the deep neural networks into the reinforcement learning has been provided, and deep reinforcement learning has become a remarkable tool to handle the complex problems. Therefore, we introduce the DQN to optimize the UAV navigation issue.

In the proposed DQN framework, since we assume there are 32 UAVs in the UAV system, the input layer is a $32 \times 32 \times 4$ space and the first hidden layer is a convolutional (conv.) layer with 8 $4 \times 4$ filters with stride 2. Followed by a rectifier nonlinear operation, the second hidden layer is designed as a conv. layer with 16 of $2 \times 2$ filters with stride 2, which reduces the dimension for suppressing the complexity without losing important information of the network. Then, the next layer is also a conv. layer with 16 filters, and the dimension of these filters is $3 \times 3$ with stride 1. The remaining hidden layer is a fully-connected (FC) layer with 256 neurons. Additionally, the output layer is a FC layer which provides the valid actions in the UAV navigation optimization.

\subsection{Learning Policy}

To enable the UAV navigation, a novel learning policy is proposed based on the developed DQN. At first, the state space $S$ is supposed to represent the received signal strengths, and this set is formulated as $S = \{ P_{R}^k < -120dBm, -120dBm \leq P_{R}^k \leq -40dBm, P_{R}^k \geq -40dBm | \forall k \}$.
Following the state space $S$, we assume $R$, $P$, and $V$ as the mean value of the immediate reward, the transition probability, and the utility function, respectively, and the $Q$-function is expressed by
\begin{align}
\label{h4}
Q^{\pi}(s, a) = R(s, a) + \tau \sum_{s^{'}\in S} P_{s s^{'}} V^{\pi}(s^{'}),
\end{align}
where $\pi$ is denoted as the policy, and our goal is to obtain the best policy $\pi^{\ast}$. Also, $s$ and $a$ represent the state and action, respectively. Concretely, the action $a$ is performed through the environment simulator, and it updates its state and its reward based on the information from the BS. Furthermore, $\tau$ defines the discount factor in the region $0 < \tau < 1$, while $S$ represents the state space. The future reward function obtained at time slot $t$ after learning the channel state over the last $T$ time slots duration is expressed as
\begin{align}
\label{h20}
R(s, a; t) = \sum_{t_0 = 0}^T \tau(t_0) r(t - t_0),
\end{align}
Here, $r(t)$ is the immediate reward function, while $t$ is defined as the time index. It is pointed out that $r(t)$ is determined by the SINR of the UAVs, which can be collected from the received signals of the UAVs. The SINR varies when the UAV moves from one position to another position in different time slot, and $r(t)$ is updated as the SINR changes. Eq. (\ref{h20}) is the sum of the rewards in different time slots and it is used to update the $Q$ state. Supposing $\alpha$ and $\eta_0$ as the positive constant and the power threshold, we formulate $r(t)$ as
$$
r(t) = \left\{
\begin{array}{rcl}
\alpha \eta_k(t),      &      & {\eta_k(t) > \eta_0,}\\
-1,      &      & {\eta_k(t) \leq \eta_0.}
\end{array}
\right.
$$

\begin{figure}
  \centering
  \includegraphics[width=87mm]{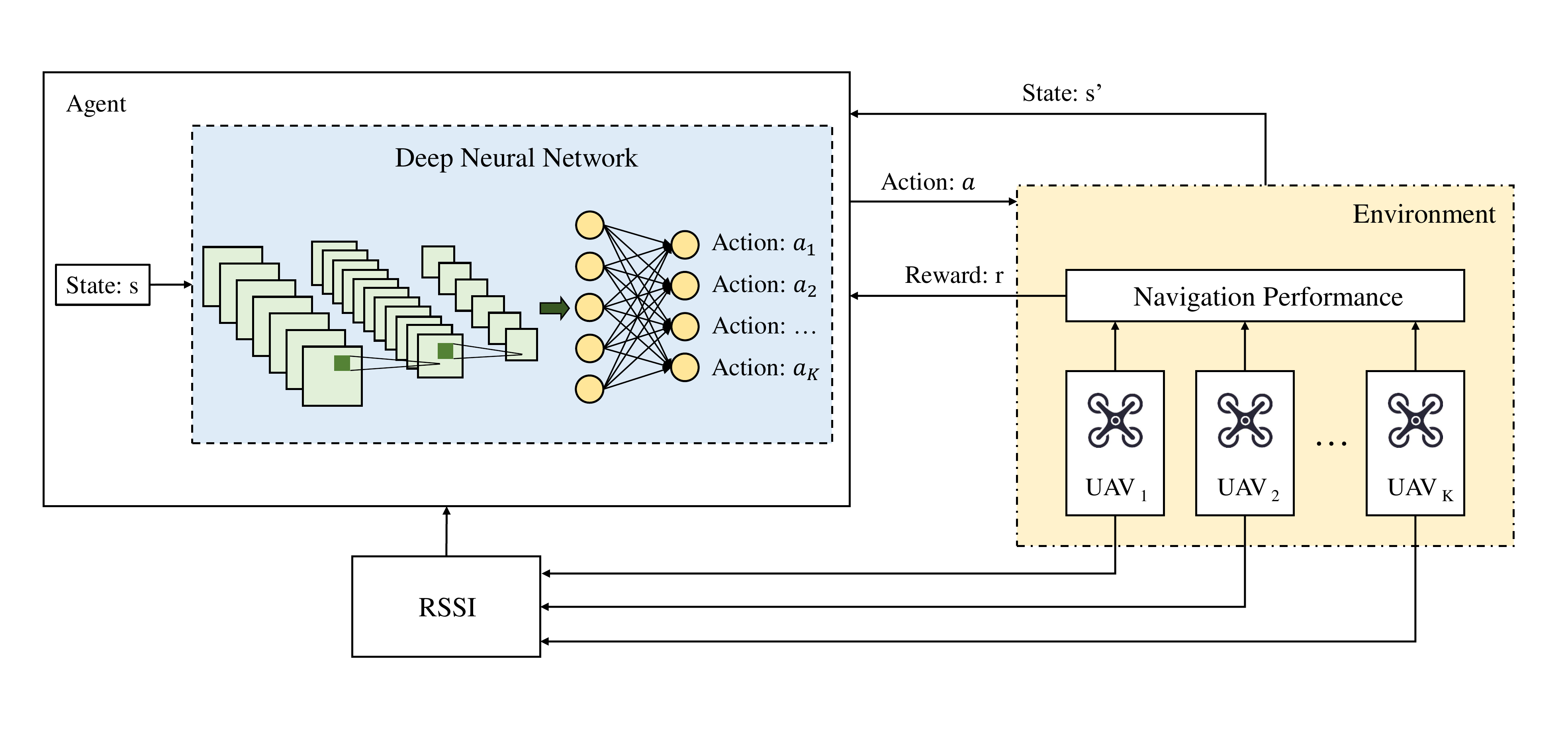}\\
  \caption{A DQN-based UAV navigation framework. }
  \label{fig_rl}
\end{figure}

Then, we obtain the maximum $Q$-function as
\begin{align}
\label{h5}
Q^{\pi^{\ast}}(s, a) & = R(s, a) + \tau \sum_{s^{'}\in S} P_{s s^{'}} V^{\pi^{\ast}}(s^{'})
\notag\\ & = \mathbb{E}[r + \tau \max_{a^{'}} Q^{\pi^{\ast}}(s^{'}, a^{'})|s, a],
\end{align}

\begin{algorithm}
   \renewcommand{\algorithmicrequire}{\textbf{Input:}}
   \renewcommand{\algorithmicensure}{\textbf{Output:}}
   \caption{DQN-Based Training Method UAV Navigation.}
   \label{alg:1}
   \begin{algorithmic}[1]
      \STATE Initialize the $Q(s, a; \omega)$ table with weights $\omega$ randomly.
      \STATE Initialize the target DQN parameters with $\omega^{-} = \omega$.
      \STATE Initialize the replay memory $M$.
      \STATE Construct the DQN structure.
      \STATE Start the environment simulator.
      \STATE \textbf{For} episode = $1, 2,\cdot \cdot \cdot, num$ \textbf{do:}
      \STATE Initialize all the beginning stages $s$ as zero.
      \STATE \textbf{For} $t = 1, 2,\cdot \cdot \cdot, T$ \textbf{do:}
      \STATE Obtain the received signal strengths at the UAVs.
      \STATE Obtain the instant reward $r(t)$ based on the received signal strengths.
      \STATE Choose an action based on the given probability $\varepsilon$.
      \STATE Observe the instant reward r(t) and the next state $s(t+1)$.
      \STATE Save the knowledges $(s(t), s(t+1), r(t), r(t+1))$ in $M$.
      \STATE Sample mini-batch of examples from the $M$ randomly.
      \STATE Adopt the stochastic gradient descent (SGD) mean to train the DQN according to Eq. (\ref{h10}).
      \STATE Update network parameters $\omega$ of the DQN.
      \STATE Update the $Q(s, a; \omega)$ table.
      \STATE \textbf{End for}
      \STATE \textbf{End for}
   \end{algorithmic}
\end{algorithm}

\begin{algorithm}
   \renewcommand{\algorithmicrequire}{\textbf{Input:}}
   \renewcommand{\algorithmicensure}{\textbf{Output:}}
   \caption{DQN-Based Testing Method for UAV Navigation.}
   \label{alg:2}
   \begin{algorithmic}[1]
      \STATE Load the DQN framework.
      \STATE Start the environment simulator.
      \STATE \textbf{Loop}
      \STATE Receive $\eta_k(t)$ of each UAV at a time slot.
      \STATE Select an action with the largest $Q(s, a; \omega)$ value in the $Q(s, a; \omega)$ table.
      \STATE Update the location of each UAV according to the actions.
      \STATE Update the environment simulator.
      \STATE Update the target values of the DQN.
   \end{algorithmic}
\end{algorithm}

Afterwards, noting that $A$ as the action space, the discounted cumulative state function is formulated as
\begin{align}
\label{h7}
V^{\pi^{\ast}}(s) = \max_{a\in A}[ Q^{\pi^{\ast}}(s, a)].
\end{align}
After obtaining the maximum $Q$-function, we need to derive the optimal policy. Using the recursive mechanism, the $Q$-function can be updated as
\begin{align}
\label{h8}
Q_{t+1}(s, a) = & Q_{t}(s, a)
\notag\\ & + \beta(r + \tau[\max_{a^{'}} Q_t(s^{'}, a^{'})] - Q_{t}(s, a)),
\end{align}
where $\beta$ defines the learning rate. Since the received signal strength fluctuates as the UAVs' position changes, $\beta$ is required to vary from different position. For example, to collect the received signal strengths when the UAV is close to the destination, the learning rate should be increased.

Thereafter, supposing $\omega_j$ as the weight at the $j$-th iteration of the DQN, the target values of the DQN can be given as
\begin{align}
\label{h9}
y = r + \tau \max_{a^{'}} Q_t(s^{'}, a^{'}; \omega_j),
\end{align}

Afterwards, to find the optimum solution, the loss function of the DQN can be designed as
\begin{align}
\label{h10}
\tu{loss}(\omega) = \mathbb{E}[(y - Q(s, a; \omega))^2].
\end{align}

After deriving the learning policy, it is noted that action selection and execution for the agents should be processed and we propose a $\varepsilon$-greedy-based policy for selecting behavior distribution. To be specific, $\varepsilon$ is denoted as the exploration probability. We select the behavior distribution which follows the greedy strategy with probability $1 - \varepsilon$ and choose an action with the biggest $Q$ value. In order to explain the proposed DQN-based navigation scheme clearly, the DQN-based navigation framework is illustrated in Fig. \ref{fig_rl}. Concretely, the proposed DQN-based strategy is provided in Algorithm \ref{alg:1} and Algorithm \ref{alg:2}.

\begin{figure*}[htbp]
  \centering
  \subfigure[Performance comparison of coverage score via coverage range of several methods.]{
  \includegraphics[width=50mm]{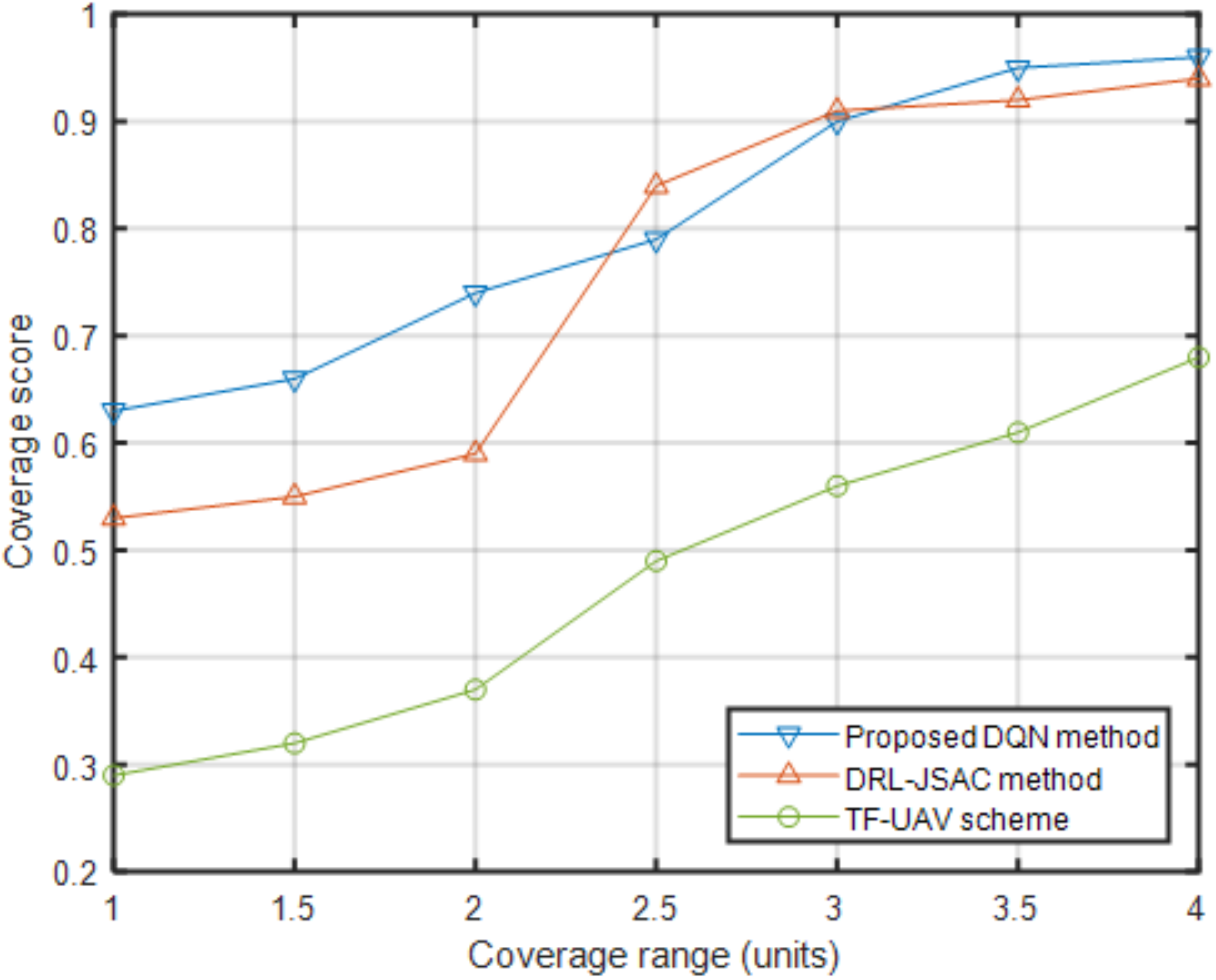}
  \label{fig_co1}
  }
  \quad
  \subfigure[Accumulated reward via epoch of the proposed DQN scheme.]{
  \includegraphics[width=50mm]{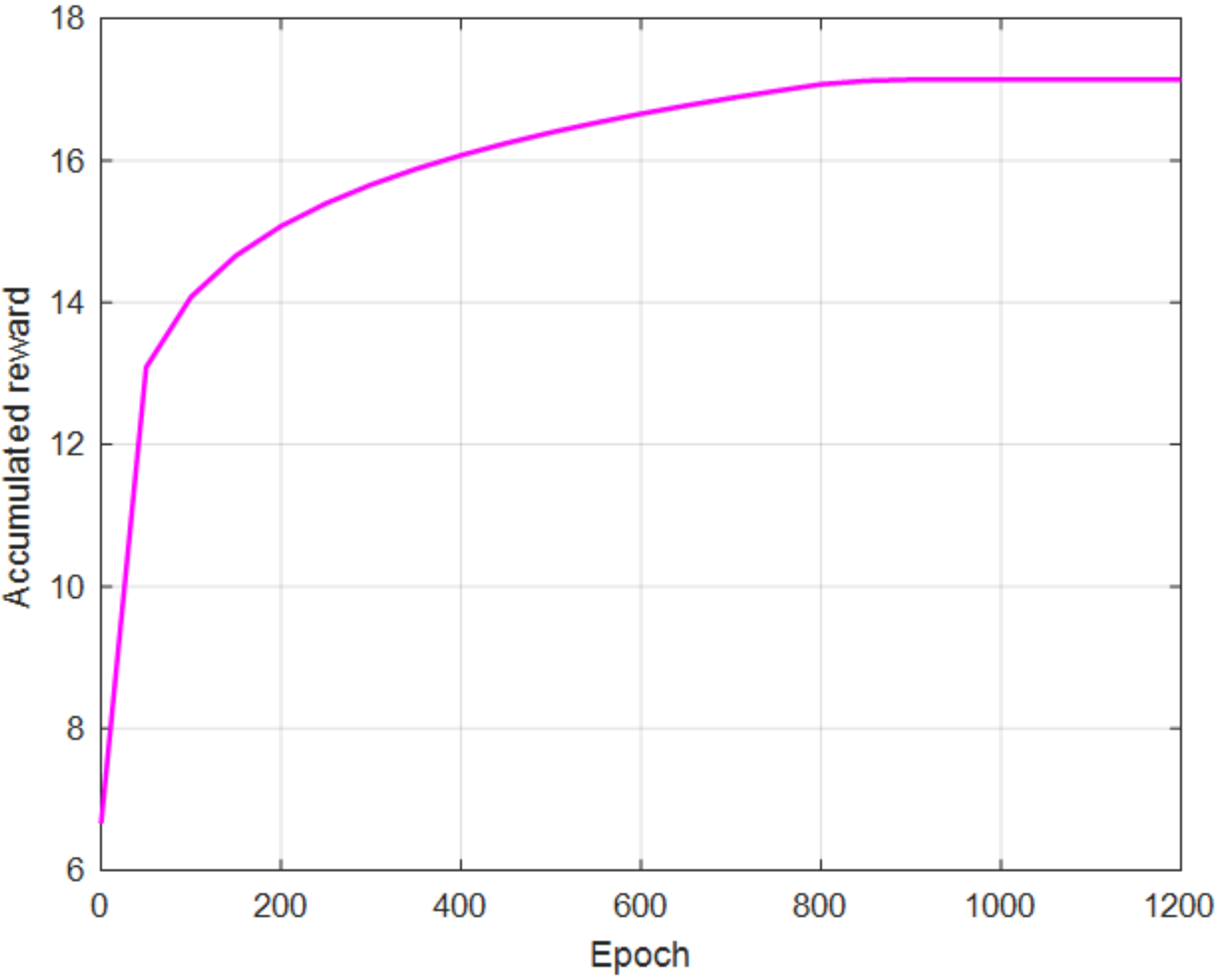}
  \label{fig_co2}
  }
  \quad
  \subfigure[UAV navigation performance.]{
  \includegraphics[width=50mm]{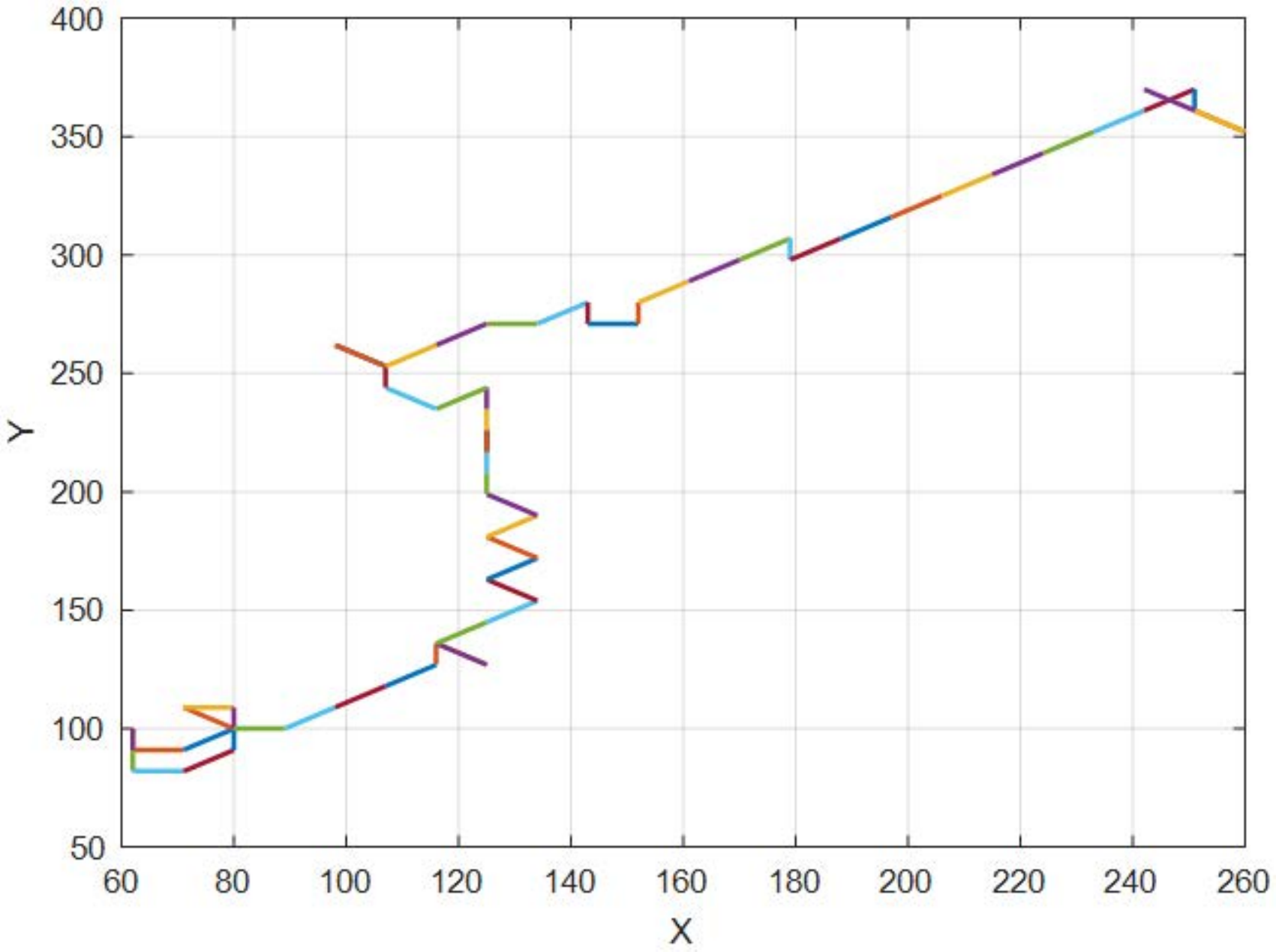}
  \label{fig_mse1}
  }
  \caption{Coverage performance of the proposed scheme.}
\end{figure*}

As described in Algorithm \ref{alg:1}, at first (Lines 1-4), we initialize the network parameters randomly. To enhance the learning stability, we introduce the target DQN and it has the same structure as the original network. Then, exploration process is conducted. The action is derived from current DQN and the action is mixed noise with Gaussian distribution to maintain the exploration. The DQN employs the SINR to update the reward function, since the SINR is regarded as the received signal strengths of the UAVs and it can reflect current location of the UAVs. This is because that the channel conditions are changing in different location and the SINR is a common index to illustrate the channel conditions. By trying all the actions for obtaining better rewards estimation, the UAVs will choose an action with highest utility (i.e., highest reward) and fly along this direction. Next (Lines 14-17), we use the mini-batch method to randomly collect examples from the replay memory. And we update the weights and bias of the network by training the DQN according to the loss function (\ref{h10}). Once the UAVs arrive at the terminal destination, the training process is stopped and the UAVs stop choosing actions.

\section{Simulation Results and Analysis}

In this section, we present numerical results of the proposed DQN-based UAV navigation scheme through massive MIMO. In our experiment, we consider a massive MIMO system with $N_t = 128$ transmit antennas and $K = 32$ single-antenna UAVs. Here, $l$ is the distance between UAV and BS. Each $\gamma^k$ is expressed as $\sqrt{\frac{\kappa}{\kappa + 1}l^{-\beta}}$ and the LoS phases follow uniform distribution over $\{-\pi, \pi\}$ radians, in which $\beta = 3.8$ and $\kappa = 6$dB are the passloss exponent and the Rician factor, respectively. Based on the Rician fading with a Rician factor of 6dB, the UAV is simulated in a $500\tu{m} \times 500\tu{m}$ indoor space. Specifically, BSs are placed at the opposite corners and  UAVs are directly above each BS, which is the worst case. The UAVs can only fly along several particular routes without running into the walls, and they need to try other directions if any crash occurs. Also, we set the total transmitted power as 20 W and the sampling period is initialized as 0.02 ms, while $d = \frac{\lambda}{2}$ is initialized. Furthermore, the batch size is 100 and the number of training examples is 250000, while the amount of testing examples is set as 50000.

To evaluate the coverage performance of the UAVs, we divide the whole area into $M$ zones, and each zone should be covered by at least one UAV at each time period. Assuming $F_m$ as the amount of time slots when the $m$-th zone is covered, the coverage score of zone $m$ is defined as
\begin{align}
\label{h15}
r_m = \frac{F_m}{T},
\end{align}

Our objective is to maximize the coverage score. In Fig. \ref{fig_co1}, we compare the coverage score via coverage range performance with that of the TF-UAV scheme \cite{tfu}, and the DRL-JSAC method \cite{tang}. It can be seen that the proposed DQN method outperforms other methods in terms of the coverage performance when the UAV coverage range is less than 2.4 or larger than 3.1. Although the DRL-JSAC method obtains the highest coverage score when the coverage range is from 2.4 to 3.1, its curve is not smooth and this result indicates that the DRL-JSAC method results in randomness and this method is not robust. Different from the result of the DRL-JSAC method, we observe that the curve of the proposed DQN-based method is smoother compared with that of other schemes, showing that the proposed method is more robust in UAV navigation. As the coverage range increases, the proposed scheme still performs very well since the curve is monotonically increasing. As can be observed in Fig. \ref{fig_co2}, the accumulated reward increases monotonically as the epoch increases. The curve grows slowly as the epoch is more than 300. It is because the zones were not well covered at the initial time and the action selection brings an improving reward. When the zones are well covered, the reward is increasing smoothly and slowly. Hence, the proposed scheme achieves the better coverage performance.

Fig. \ref{fig_mse1} shows the navigation performance of the DQN-based scheme. It can be seen from Fig.~\ref{fig_mse1} that the UAV is moving away from the origin of the cartesian coordinate system as the time increases, which indicates that the UAV can fly under accurate UAV navigation. Also, we observe that part of the navigation curve changes sharply, which are induced by the fact that the UAV encounters obstacles such as walls of the indoor space during flying, which implies that the proposed DQN-based method is capable of extracting the environment information and making the best decision.

\begin{figure*}[htbp]
  \centering
  \subfigure[Performance comparison of several typical schemes in terms of convergence performance.]{
  \includegraphics[width=50mm]{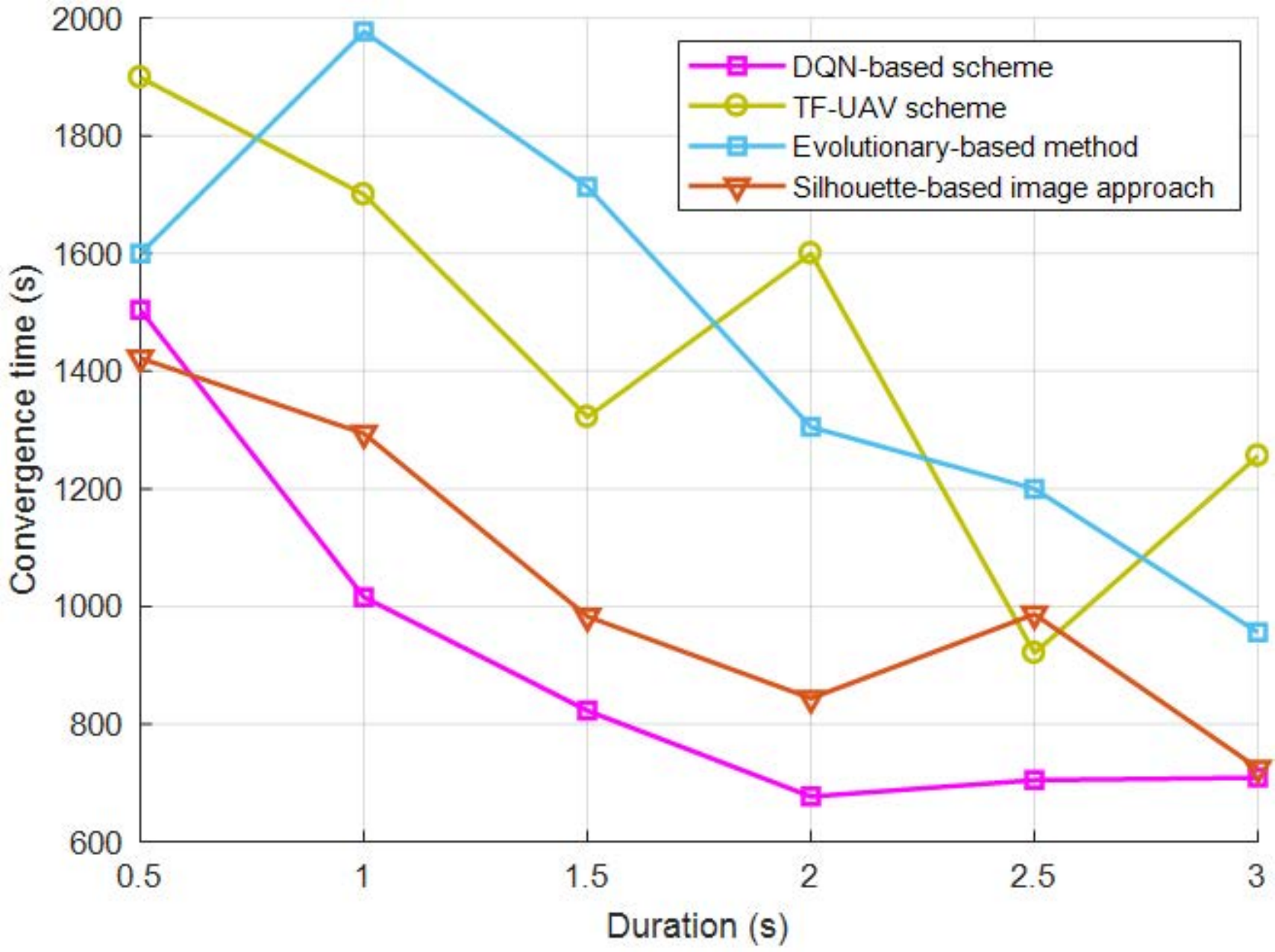}
  \label{fig_mse2}
  }
  \quad
  \subfigure[Learning rate.]{
  \includegraphics[width=50mm]{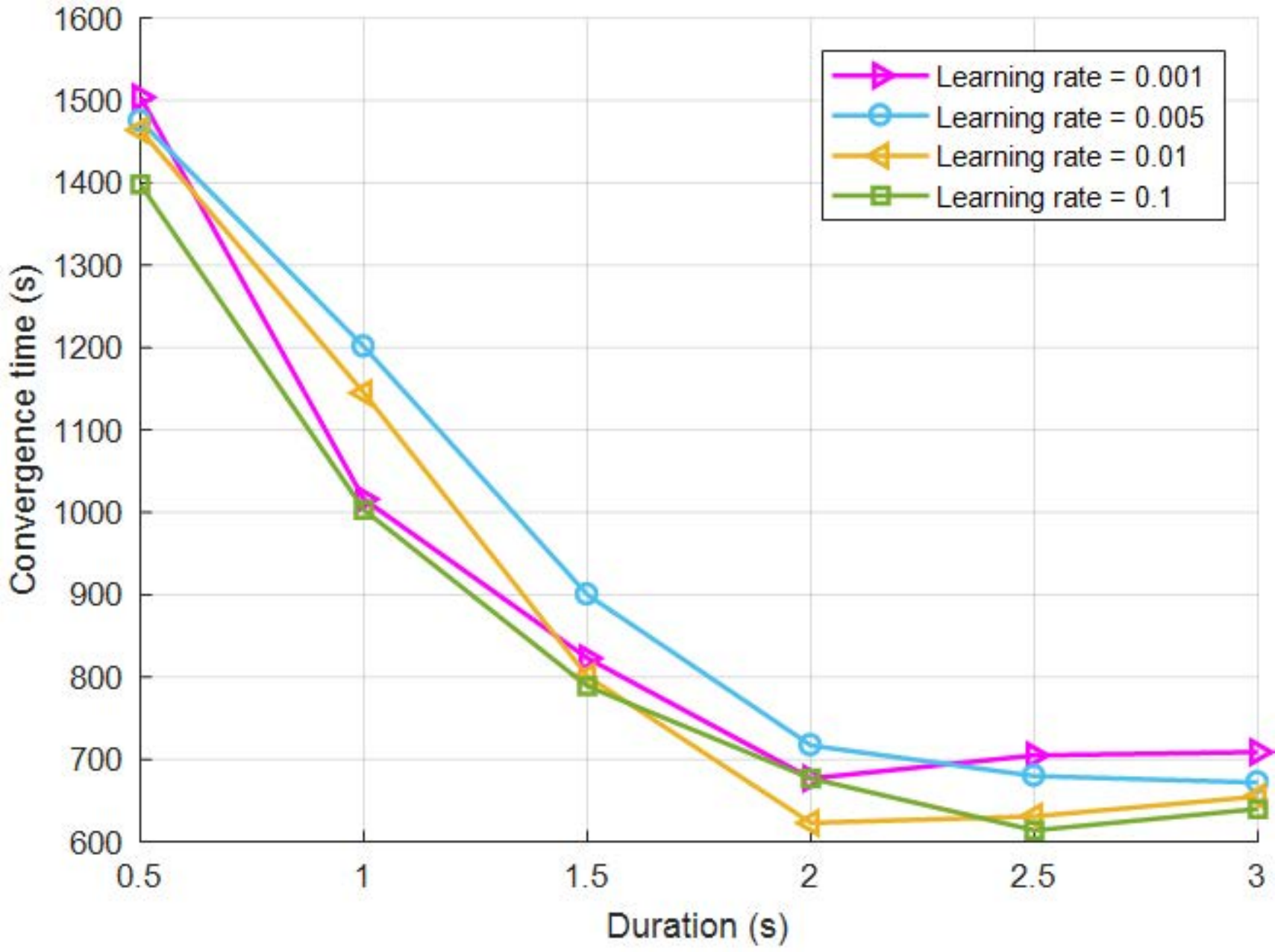}
  \label{fig_mse3}
  }
  \quad
  \subfigure[UAV speed.]{
  \includegraphics[width=50mm]{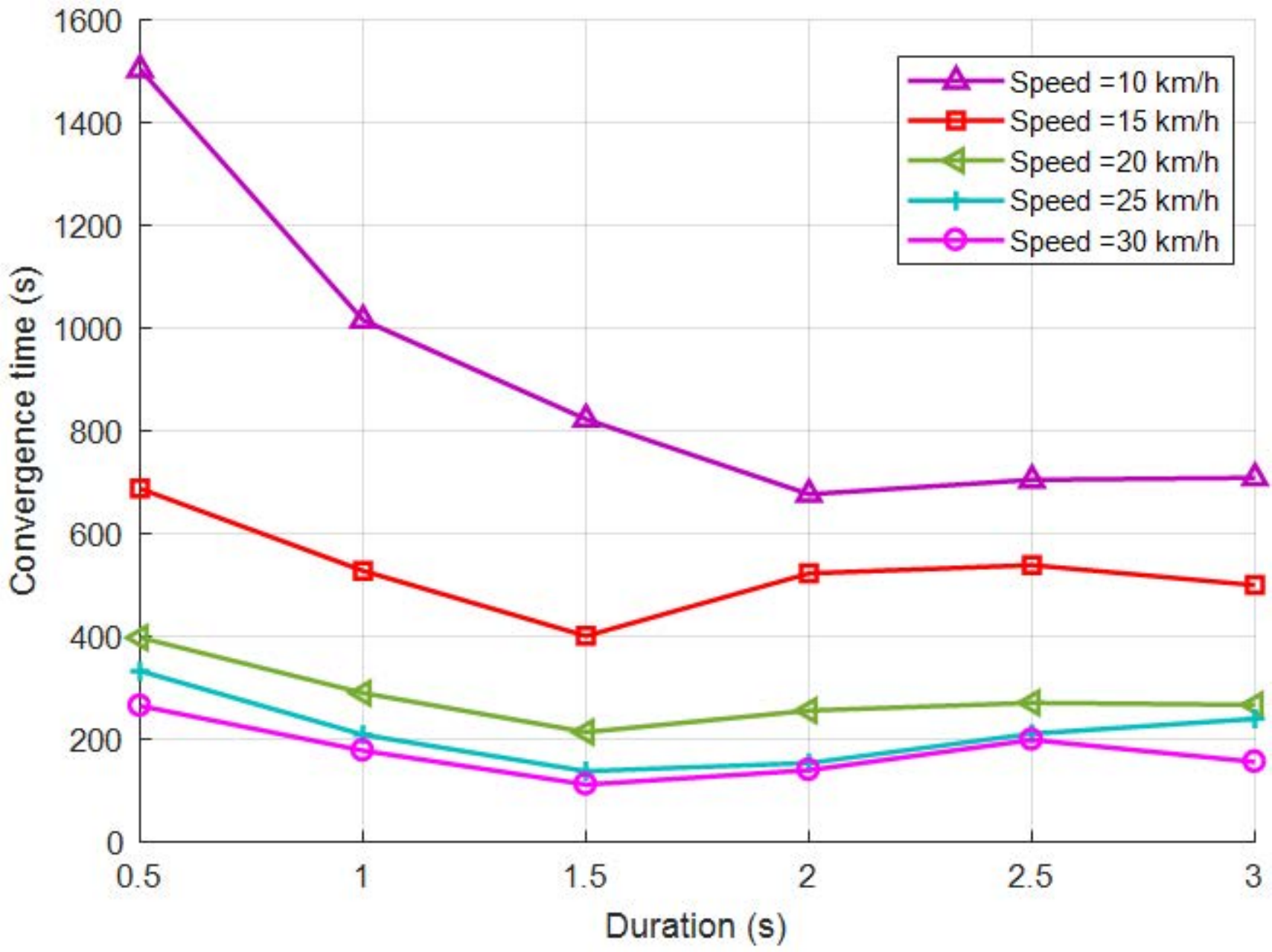}
  \label{fig_mse4}
  }
  \caption{Convergence performance of the proposed scheme.}
\end{figure*}

The performance comparison of the convergence time of the UAV navigation against the sampling duration is presented in Fig. \ref{fig_mse2}, in which the DQN-based scheme, the TF-UAV scheme \cite{tfu}, the evolutionary-based method \cite{ev1}, and the Silhouette-based image approach \cite{sil} are included. Here, the speed of the UAV is set as 10 km/h. It is observed from Fig.~\ref{fig_mse2} that the convergence time of each UAV navigation algorithm is reduced as the sampling duration increases, for the reason that the probability of making a wrong decision increases as the UAVs travel a longer distance. Meanwhile, it can be seen that when the sampling duration approaches a threshold time, the speed of the convergence of all the algorithms reduces and the speed would reduce to 0 in theory, which means that the algorithm converges. In particular, the proposed DQN-based scheme converges when the sampling duration is 2s, while other methods still shake sharply. Also, the proposed scheme requires less convergence time compared with that of other schemes in most cases, although the Silhouette-based image approach requires less convergence time when the sampling duration increases from 0.5s to 0.6s.

Fig. \ref{fig_mse3} exhibits the convergence performance of the UAV navigation against the sampling duration of the DQN-based method, where the initial learning rate is set as 0.1, 0.01, 0.005, and 0.001, respectively. Initially, the speed of the UAV is 10 km/h. Learning rate is an essential parameter in a deep learning-based approach, which is always introduced to evaluate the robustness and convergence performance of a deep learning-based method. We observe from Fig.~\ref{fig_mse3} that the DQN-based scheme converges quickly when adopting a larger initial learning rate, due to the fact that a larger initial learning rate facilitates the convergence behavior. However, it should be noted that a larger learning rate leads to a strong vibration and degrades the convergence performance. As shown in Fig.~\ref{fig_mse3}, the curve is more stable and it becomes smooth finally when introducing the learning rate as 0.001. It indicates that a smaller learning rate enhances the UAV navigation performance.

Fig. \ref{fig_mse4} shows the convergence performance of the UAV navigation when the UAV flies at different velocities, in the case of 10 km/h, 15 km/h, 20 km/h, 25 km/h, and 30 km/h, respectively. Here, the initial learning rate is set as 0.001. It is observed from Fig.~\ref{fig_mse4} that a larger velocity requires a smaller convergence time compared to that at a smaller velocity, since a larger velocity reduces the responsiveness of the UAV system. However, it can also be seen from Fig.~\ref{fig_mse4} that the tendency of the curves is that they are not decreasing in general and they shake frequently when adopting a larger UAV speed, which indicates that a larger UAV speed degrades the UAV navigation performance.

\balance

\section{Conclusions}
\label{secconclusion}

In this paper, we have presented a deep reinforcement learning-based scheme for UAV navigation through massive MIMO. Specifically, we first design an efficient DQN which comprises conv. layers and FC layers to extract useful features of the massive MIMO. In addition, a Q-learning-based learning policy is proposed to realize the UAV navigation. Here, we treat each UAV-ground link as an agent, and the optimal location at the UAVs is obtained based on the received signal strengths without requiring global information. Numerical results also show the superior UAV navigation performance of the DQN-based strategy compared with several typical strategies in terms of convergence and coverage.

%Fang
%Yang

%-------------------

\end{document}